\documentclass[pra,aps,superscriptaddress,twocolumn,showpacs,longbibliography,amsmath,amssymb]{revtex4-2}

\usepackage{graphicx}
\usepackage{dcolumn}
\usepackage{bm}
\usepackage{stackrel}
\usepackage{color}

\begin{document}


\title{Evidence of 1+1D photorefractive stripe solitons deep in the Kerr limit}

\author{Ludovica Falsi}
\affiliation{Dipartimento di Fisica, Universit\`a di Roma “La Sapienza”, 00185 Rome, Italy}
\email{ludovica.falsi@uniroma1.it}
\author{Alberto Villois}
\affiliation{Dipartimento di Fisica, Universit\`a degli Studi di Torino, 10125 Torino, Italy}
\author{Francesco Coppini}
\affiliation{Dipartimento di Fisica, Universit\`a di Roma “La Sapienza”, 00185 Rome, Italy}
\author{Aharon J. Agranat}
\affiliation{The Institute of Applied Physics, The Hebrew University, Jerusalem 91904, Israel}
\author{Eugenio DelRe}
\affiliation{Dipartimento di Fisica, Universit\`a di Roma “La Sapienza”, 00185 Rome, Italy}
\author{Stefano Trillo} 
\affiliation{Department of Engineering, University of Ferrara, 44122 Ferrara, Italy}


\date{\today}

\begin{abstract} 
\noindent The Kerr nonlinearity allows for exact analytic soliton solutions in 1+1D. While nothing excludes that these solitons form in naturally-occurring real-world 3D settings as solitary walls or stripes, their observation has previously been considered unfeasible because of the strong transverse instability intrinsic to the extended nonlinear perturbation.  We report the observation of solitons that are fully compatible with the 1+1D Kerr paradigm limit hosted in a 2+1D system. The waves are stripe spatial solitons in bulk copper doped potassium-lithium-tantalate-niobate (KLTN) supported by the unsaturated photorefractive screening nonlinearity. The parameters of the stripe solitons fit well, in the whole existence domain, with the 1+1D existence curve that we derive for the first time in closed form starting from the saturable model of propagation. Transverse instability, that accompanies the solitons embedded in the 3D system, is found to have a gain length much longer than the crystal. Findings establish our system as a versatile platform for investigating exact soliton solutions in bulk settings and in exploring the role of dimensionality at the transition from integrable to non-integrable regimes of propagation.
\end{abstract}

\maketitle

{\bf Introduction.} Dimensionality plays a crucial role in nonlinear wave physics. While in statistical mechanics \cite{Landau2013} and string theory \cite{String2004}, higher dimensions are key to formulating consistent models, in nonlinear waves, it is lower dimensions that support interesting effects, such as integrability \cite{Drazin1989}. Low-dimensional paradigms of this sort can be found in many fields, 
with notable examples including the transport properties of 2D metals in the Quantum Hall Effect \cite{Halperin2020}, anomalous thermal conduction in 1D nanowires \cite{Dematteis2020}, and the 2D BKT transition in condensed matter systems \cite{KTbook}. A basic example of low-dimensional integrability is the Nonlinear Schr\"odinger Equation (NLSE), a paradigm model that describes a large variety of different physical systems \cite{Drazin1989, Whitham1974, Kivshar2003}, such as for picosecond optical pulses in fiber \cite{Mollenauer1980}, photonic crystals and waveguides \cite{Dudley2006}, 
and matter-waves in Bose-Einstein condensates \cite{Cornish2020}. Here, the 1+1D NLSE supports solitons with an explicit analytical expressions that behave like solitary particles, unable to exchange energy. The question naturally arises as to what these low-dimensional paradigms have to say about real-world 3D settings, where the 1+1D soliton should form as an extended solitary wall or stripe. For example, does integrability cross over to stripe Kerr solitons? To date, no experimental evidence of unbroken 1+1D Kerr solitons hosted in a 2+1D homogeneous system has been reported.

One widely studied system that allows the formation of 1+1D solitons hosted in a full 3D environment is that of photorefractive (PR) crystals \cite{Yeh1993,solitonbook,Chen2012}. Here, spatial optical solitons are observed in many different schemes, stable localized light beams that form out of the balance between diffraction and nonlinear self-focusing. Discovered almost three decades ago, self-trapped beams in PR crystals still play a major role both in expanding soliton phenomenology and in exploring the role of solitons in numerous different phenomena, such as rogue waves \cite{Onorato2013, Pierangeli2015, Marsal2014}, optical turbulence \cite{Picozzi2014, Pierangeli2016}, and replica-symmetry-breaking \cite{Pierangeli2017}. From the applicative point of view, PR solitons allow the imprinting of electro-optic integrated circuits, such as highly miniaturized fast switching and routing devices for all-optical information processing and transparent communication systems \cite{Ercole2004, Asaro2005, Lan1999}. The PR nonlinearity is the macroscopic result of different processes, i.e., the formation of a space-charge field caused by the diffusion and drift of photogenerated charges, and crystal electro-optic response, the resulting model being system-dependent  \cite{solitonbook, DelRe09}. For example, in unbiased lithium-niobate LiNbO$_3$, it is dominated by the bulk photovoltaic response, this leading to a self-defocusing nonlinearity able to support dark solitons \cite{Taya1995}. In unbiased near-transition potassium-lithium-tantalate-niobate KTa$_{1-x}$Nb$_x$O$_3$:Li (KLTN), the nonlinearity is governed by charge diffusion, leading to scale-free optical propagation \cite{DelRe2011}. In most PR crystals, this including strontium-niobate Sr$_x$Ba$_{1-x}$Nb$_2$O$_3$ and KLTN, there is an experimentally accessible regime in which charge drift is made dominant through the application of an external bias electric field, the result being what are termed PR screening solitons, the most widely studied type of spatial soliton \cite{DelRe98, DelRe03, DelRe09}. The leading model for screening solitons is a saturated Kerr-like nonlinearity. In unsaturated conditions, the screening nonlinearity becomes the Kerr nonlinearity, so that in 1+1D, laser stripes (in 3D) should be governed by the NLSE. While 1+1D PR stripe solitons in nonintegrable saturated conditions are commonly observed \cite{Yeh1993, Segev1992}, soliton stripes in the Kerr-limit have not been reported.  These have been previously considered experimentally inaccessible because of the effects of strong transverse instability in regimes where integrable models hold \cite{Stageman1999}.  According to this commonly held idea, a Kerr stripe soliton will necessarily break-up into an array of spots \cite{TI, Mamaev1996, Zakharov1974}, an instability intrinsic to the fact that the system hosting the 1+1D soliton is 3D. Short of limiting experiments to PR waveguides \cite{Aitchison, Maneuf1988} or photonic lattices \cite{Yang2012}, or introducing fundamentally different waves, such as incoherent Kerr solitons \cite{Anastassiou2000, Anderson2004}, PR Kerr solitons in the bulk have commonly be considered to be beyond experimental reach. 

We here experimentally demonstrate, for the first time, Kerr 1+1D stripe photorefractive solitons in bulk paraelectric KTN:Li.  
In conditions in which beams are governed by the PR screening nonlinearity, a Kerr-saturated response supported by the quadratic electro-optic, the Kerr regime is demonstrated by comparing observed PR soliton existence parameters with those predicted for the screening nonlinearity, that we here predict analytically. Strong agreement with the paradigm Kerr nonlinearity is found for the highly unsaturated self-trapping regime and for propagation distances such that transverse instability build-up, here analyzed both experimentally and numerically, remains negligible.

{\bf Theory.} The paraxial nonlinear propagation in bulk paraelectric KTN:Li is described by the dimensionless model \cite{DelRe98,DelRe09}

\begin{equation} \label{satNLS}
{\displaystyle i \frac{\partial \psi}{\partial z} + \nabla^2 \psi  -  \frac{\psi}{(1+ |\psi|^2)^2} =0},
\end{equation}

\noindent where $\psi(x,y,z)=E(X,Y,Z)/\sqrt{I_b}$ is the normalized slowly-varying optical field in units of background intensity illumination $I_b$, 
$\nabla^2 \equiv \partial_x^2 + \partial_y^2$, with $(x,y)$ and $z$ being transverse and longitudinal variables with corresponding real-world variables $(X,Y)=X_0 (x,y)$ and $Z=Z_{nl} z$, where $X_0= \left[ 2k_0 n_1 \chi \right]^{-1/2}$ and $Z_{nl}=\chi^{-1}$ are transverse and longitudinal characterstic lengths, respectively.  Here $k_0$ is the vacuum wavenumber, $n_1$ the unperturbed refractive index, and $\chi= \frac{1}{2} k_0 n_1^3 |g_{eff}| \epsilon_0^2 (\epsilon_r(0)-1)^2 E_0^2$, with $g_{eff}$ the effective quadratic electro-optic coefficient, $\epsilon_0$ the vacuum dielectric constant, $\epsilon_r(0)$ the low-frequency relative dielectric constant, and $E_0$ the static bias electric field applied to the crystal.

We are interested in soliton stripe solutions of Eq. \eqref{satNLS} confined along $x$. By substituting the ansatz $\psi(x,y,z)=u(x) \exp(-i \gamma z)$ in Eq. \eqref{satNLS}, where $\gamma$ is the soliton nonlinear phase shift or eigenvalue, the soliton profile $u(x)$ is found to obey the eigenvalue equation ($\dot{u} \equiv du/dx$): 
\begin{equation} \label{soleig}
{\displaystyle \ddot{u} -  \frac{u}{(1+ u^2)^2} + \gamma u = 0},
\end{equation} 
\noindent which is equivalent to the Newton equation $\ddot{u}=-\partial V(u)/\partial u$ for the 1D motion along coordinate $u$ of an ideal unit mass particle in a potential well $V(u) = \frac{\gamma u^2(1 + u^2) + 1}{2(1+u^2)}$. Equation \eqref{soleig} implies conservation of energy (Hamiltonian) $E=H(u,\dot{u})=\dot{u}^2/2 + V(u)$, which we exploit to obtain the existence curve in analytical form as follows. 
Solitons correspond to the separatrix trajectories of Eq. \eqref{soleig}, namely orbits that asymptotically connect the saddle point $u=\dot{u}=0$ of the Hamiltonian with itself at $x=\pm \infty$, and hence have fixed energy $E=H(0,0)=1/2$. The soliton peak amplitude $u=u_0$ is given by the maximum elongation along these orbits, which is reached at the return point of the potential where $\dot{u}=0$. The constraint $V(u=0)=V(u=u_0)$ allows us to find the soliton peak $u_0$ as the function of the eigenvalue as
\begin{equation}\label{u0theory}
    u_0=\sqrt{\frac{1}{\gamma}-1},
\end{equation}
which entails that the soliton family covers the parameter range $0<\gamma<1$. On the other hand, the conservation of energy yields
$\dot{u}=\pm \sqrt{2(E-V(u))}=\pm \sqrt{1-2V(u)}$, which can be cast, using Eq. \eqref{u0theory}, in the form
\begin{equation} \label{1storder}
\dot{u} = \pm  \sqrt{\gamma u^2 \frac{u_0^2 - u^2}{1+u^2} }.
\end{equation}
Integration by quadrature of Eq. \eqref{1storder} yields the inverse soliton profile $x=x(u,u_0)$ in closed form:
\begin{equation} \label{xofu}
x=\gamma^{-1/2} \left[ \sin^{-1} g(u^2) + u_0^{-1} \tanh^{-1} h(u^2) \right],
\end{equation}
where $g(u^2)\equiv \sqrt{\frac{u_0^2 - u^2}{1+u_0^2}}$ and $h(u^2) \equiv \frac{1}{u_0} \sqrt{\frac{u_0^2-u^2}{1+u^2}}$. Although Eq. \eqref{xofu} cannot be inverted to obtain the analytical soliton profile in sufficiently simple form, it allows the calculation of the intensity full-width-half-maximum (FWHM), say $\Delta x$, by recalling its implicit definition $u^2(x=\Delta x/2)=u_0^2/2$. Setting $x=\Delta x/2$,  $u^2=u_0^2/2$ in Eq. \eqref{xofu}, and using Eq. \eqref{u0theory}, we  find the explicit existence curve $\Delta x=\Delta x(u_0)$ (i.e. FWHM as a function of peak amplitude) of the whole soliton family, which reads
\begin{eqnarray} \label{existence}
\Delta x = 2\sqrt{1+u_0^2} \left[  \sin^{-1} g_0  + \frac{1}{u_0} \tanh^{-1} h_0  \right] ,
\end{eqnarray} 
where $g_0 \equiv g\left( \frac{u_0^2}{2} \right) = \sqrt{\frac{u_0^2}{2(1+u_0^2)}}$, $h_0 \equiv h\left( \frac{u_0^2}{2} \right)= \sqrt{\frac{1}{2+u_0^2}}$.

\begin{figure}[h!]
    \centering
    \includegraphics[width=\columnwidth]{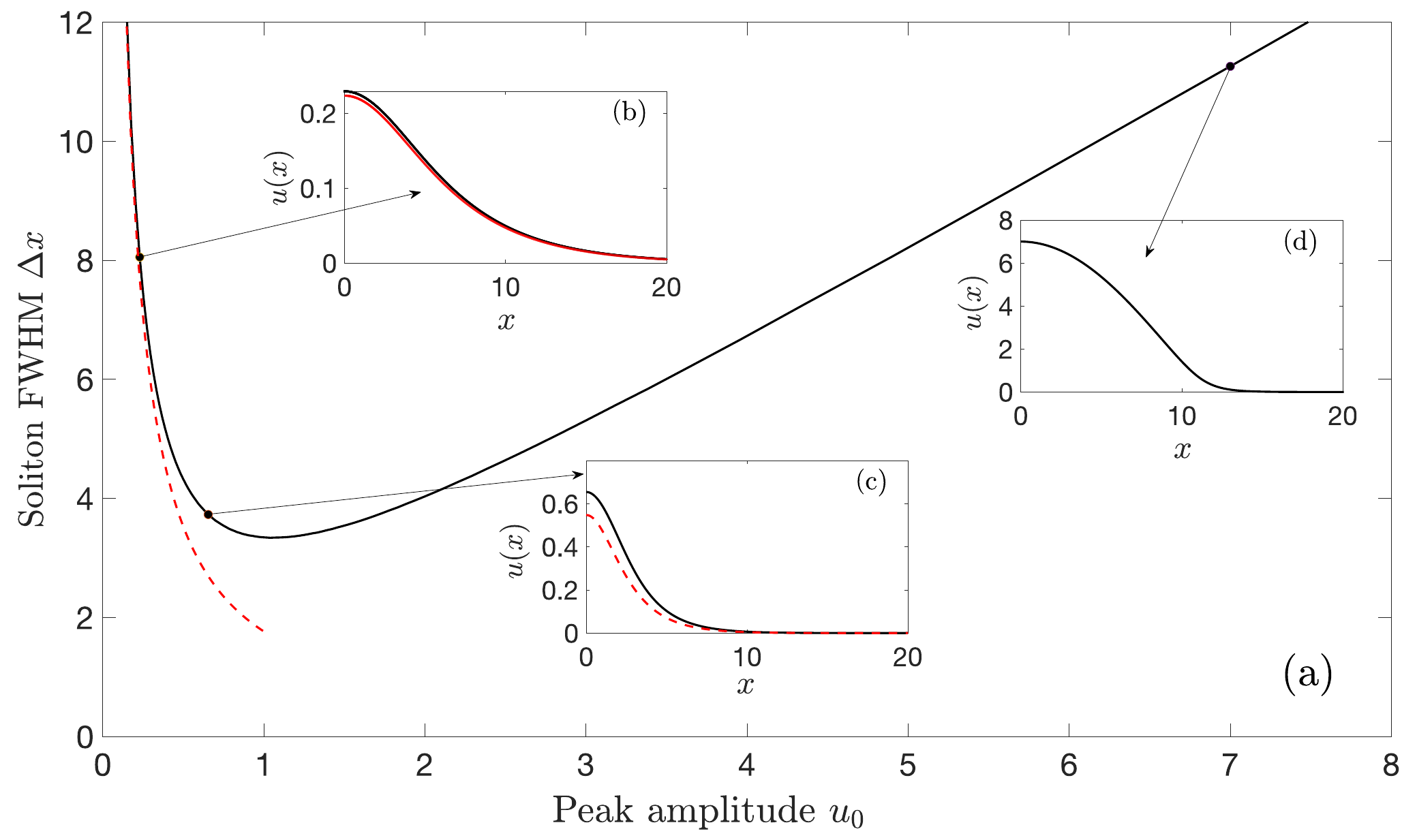}
    \caption{(a) Existence curves, FWHM $\Delta x$ vs. $u_0$ of the soliton family, comparing exact result from Eq. \eqref{existence} (solid black curve), with Kerr limit (dashed red curve). Insets: soliton profiles (exact, black solid; Kerr limit, dashed red) sampled at (b) $\gamma=0.95$, Kerr region; (c) $\gamma=0.7$, near the minimum where the Kerr limit starts to exhibit significant deviations; (d) $\gamma=0.02$, fully saturated regime. 
    }
    \label{fig:th}
\end{figure}
The existence curve Eq. \eqref{existence} is illustrated in Fig. \ref{fig:th}(a). Three distinct regimes can be discerned based on the soliton peak amplitude $u_0$. For $u_0 \gtrsim 1.5$, the system is in the fully saturated regime, while for $u_0 \sim 1$, a transition regime emerges where the soliton width goes through a minimum. For $u_0 \ll 1$, the system displays the Kerr limit, such that the approximation $(1+ u^2)^{-2} \simeq 1 - 2 u^2$ holds true.
Here, Eq. \eqref{soleig} reduces to $\ddot{u} + 2u^3 - \gamma_{K} u = 0$, characteristic of the NLSE, where the form of the effective Kerr phase-shift $\gamma_{K}\equiv (1-\gamma)>0$ reflects the fact that the refractive index change is negative but saturated in the high intensity portion, thereby giving rise to an effective self-focusing potential.
In the Kerr limit, the soliton profile has the explicit form
\begin{equation} \label{Kerrsol}
\psi(x,z)=\sqrt{1-\gamma}~{\rm sech} (\sqrt{1-\gamma}~x)~e^{-i\gamma z}.
\end{equation}
The corresponding existence curve $\Delta x= 2 \cos^{-1} \sqrt{2}/u_0 \simeq 1.76/u_0$, reported in Fig. \ref{fig:th}(a) as a dashed red curve, approximates well the exact curve [Eq. \eqref{existence}] in a large portion of the negative slope branch corresponding to eigenvalue range $\gamma \simeq [0.8,1]$. In this domain, the exact profiles [i.e. inverse of Eq. \eqref{xofu}] and their Kerr approximation [Eq. \eqref{Kerrsol}] are practically indistinguishable, as shown in Fig. \ref{fig:th}(b). Discrepancies start to appear near the minimum, though the Kerr profiles still provide, at same $\gamma$, a reasonable approximation, as shown in Fig. \ref{fig:th}(c). Clearly, over the positive slope branch, the Kerr limit breaks down.


{\bf Experiment.} Light propagation is investigated by focusing an x-polarized  Gaussian beam from a doubled 30 mW Nd:YAG laser (wavelength $\lambda=2\pi/k_0=532$ nm) onto the input xy facet of a compositionally nanodisordered photorefractive KLTN crystal using a cylindrical lens (of focal length $f=135$mm). The cylindrical lens is rotated so as to tightly focus the beam in the x-direction. Input and output transverse intensity distributions along the z axis are imaged, using a moveable spherical  lens (of focal length $50$mm), onto a CMOS camera. The zero-cut optical quality 2.1$^{(x)}$ $\times$ 1.9$^{(y)}$ $\times$ 2.5$^{(z)}$ mm KLTN crystal ($K_{0.964}$Li$_{0.036}$Ta$_{0.60}$Nb$_{0.40}$O$_3$) is grown through the top-seeded solution method and the photorefractive response for visible light is associated with deep inband Cu and V impurities. A current-controlled Peltier junction keeps the crystal  at T = 302 K (9 K above Curie temperature $T_C$) in the paraelectric phase, in conditions in which the effects of wavelength-scale index-of-refraction fluctuations are negligible while the nonlinear response is enhanced by an elevated low-frequency dielectric constant ($\varepsilon_r\simeq1 \times 10^4$). Solitons are observed when an external bias field $E_0$ is applied in the x-direction, parallel to the linear polarization of the focused beam. Saturation in the nonlinearity is fixed by  illuminating the sample from the top (in the y direction) with a uniform background intensity from a 15 mW laser at $\lambda=633$ nm.  Results refer to steady-state conditions, i.e., those for which the photorefractive nonlinear response no longer depends on time ($t\gg $10 s in our setup).  Solitons are investigated for different conditions of $u_0=\lvert E \rvert/\sqrt{I_b}$, that is, the soliton amplitude normalized to the square root of the mean value background intensity (rescaled to correctly take into account the difference in wavelengths used). For each measured value of $u_0$, we calculate the normalized width $\Delta x= \Delta X/X_0=\sqrt{2k_0 n_1 \chi} \Delta X$, where $\Delta X$ is the physical beam FWHM of the intensity profile, and $X_0$ is calculated with the KLTN unperturbed refractive index  n$_1=2.3$,  effective quadratic electro-optic coefficient $g_{eff}=0.14$m$^4$C$^{-2}$, and the field $E_0=V/L_x$ corresponding to the applied voltage $V$ (range $400\div700$ V), for each value of $u_0$, over the crystal width $L_x=$ 2.1 mm.


\begin{figure}[h!]
    \centering
    \includegraphics[width=\columnwidth]{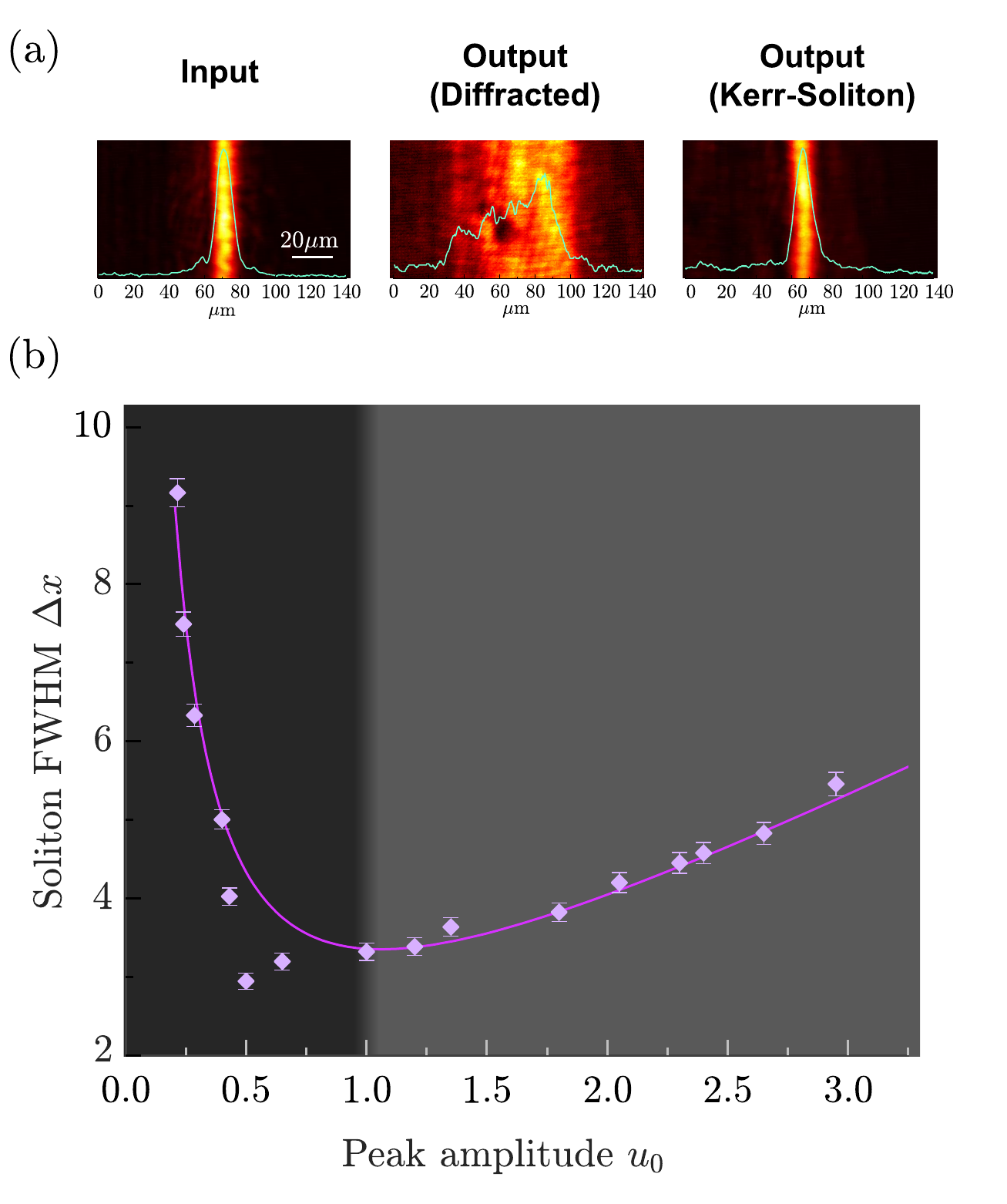}
    \caption{(a) Profiles of the input beam (left) and the diffracted output beam (middle) when nonlinearity is deactivated, and the soliton output (right) in the  Kerr limit, with a peak amplitude $u_0$ = 0.43.  (b) Soliton existence curve: experimentally measured points (diamonds) compared to the analytical curve from Eq. \eqref{existence} (solid curve).
    }
    \label{fig:exp}
\end{figure}

1+1D solitons appear as self-trapped beams whose diffraction is exactly balanced by the PR effective self-focusing nonlinearity.  An example of soliton formation for the unsaturated Kerr case of $u_0=$0.43 is reported in Fig. \ref{fig:exp}(a), where the input intensity distribution (left) is observed to diffract from 12.7 $\mu$m to 54 $\mu$m FWHM at the output in the absence of nonlinearity ($E_0=0$, center) and form a self-trapped steady-state (after 120s) non-spreading beam for $E_0=1.9$ kV/cm (right). Solitons are inspected for values of $0.21<u_0<2.95$, intensity FWHM $\Delta x=10 \mu$m and $\Delta x=12.7 \mu$m.  Corresponding values of soliton-supporting $E_0$ range from 1.9 to 3.3 kV/cm. 
Soliton normalized existence conditions from the experimental data are reported in Fig. \ref{fig:exp}(b) (diamonds) and compared to the analytical existence curve (solid line, Eq. \eqref{existence}).
The uncertainty in the existence conditions is dominated by the uncertainty in the measurements of $\Delta x$ and self-trapping $E_0$.

{\bf Discussion.}  
Findings indicate an excellent agreement between theory and experiment in both the predominantly Kerr ($u_0<0.3$) and the strongly saturated ($u_0>1$) regimes.  
An interesting discrepancy emerges for the transition region $0.43 <u_0< 0.65$.  This discrepancy may indicate the involvement of more complicated physical mechanisms at $u_0 \sim 1$. 
We attribute this to the contribution of higher-order corrections to the PR model of Eq. (1), corrections that are largest for $u_0 \sim 1$ (see, for example, Eq. (38) in Ref.~\cite{DelRe09}). 

\begin{figure}[ht]
    \centering
    \includegraphics[width=1\columnwidth]{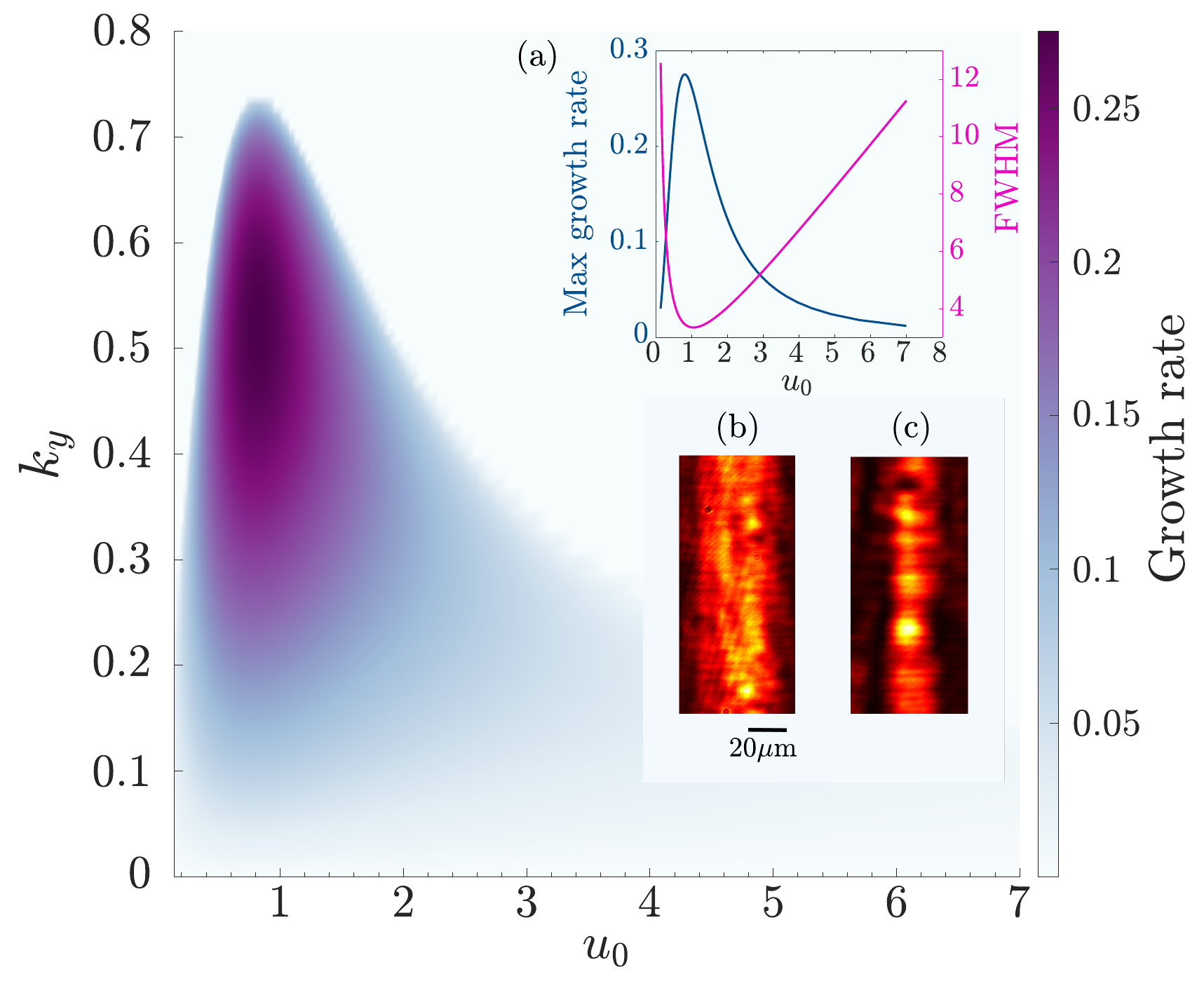}
    \caption{Pseudo-color map of TI gain $g_{TI}=g_{TI}(u_0,k_y)$ vs. soliton peak amplitude $u_0$ and transverse wavenumber $k_y$. Insets: (a) Peak gain (blue curve) vs. $u_0$ compared with the existence curve (red curve); (b,c) Output beams in $(x,y)$ plane for $u_0=0.21$, when the input FWHM is reduced to 10 $\mu$m, for two different bias showing dominant diffraction (b, \textcolor{black}{ $V=405$ V}) and break-up due to TI (c, $V=705$ V).
    }
    \label{fig:trinst}
\end{figure}



The stripe nature of the 1+1D Kerr soliton, i.e., the fact that the 1+1D nonlinear wave is embedded in a 2+1D nonlinear environment, is expected to activate transverse instabilities (TIs) \cite{Zakharov1974, Mamaev1996, TI}. In bulk, TIs can destabilize the soliton stripe against the growth of a periodic perturbation in $y$. Data reported in Fig.~\ref{fig:exp} obtained with the wider beams (FWHM of 12.7 $\mu m$) shows no evidence of transverse breakup, regardless of the choice of $u_0$ along the existence curve. To understand this point, we assess theoretically the impact of TI, by substituting in Eq.~\eqref{satNLS} the  ansatz $\psi(x,y,z)=u(x)e^{-i\gamma z}+\delta u(x) e^{-i(k_y y+\gamma z)} e^{\lambda z}+\delta v(x)^* e^{i(k_y y+\gamma z)} e^{\lambda^*z}$, where $u(x)$ is the soliton profile, and $\lambda$ is the TI growth rate. After linearization with respect to the small perturbation amplitude  $\textbf{A}=(\delta u,\delta v)^T$, we obtain and numerically solve the eigenvalue problem 
$\mathcal{M} \textbf{A}=\lambda \textbf{A}$, where
\begin{equation}\label{matrix1}
\mathcal{M}=
\begin{bmatrix}
\mathcal{L}+G(u)(1-u^2)  && -2 G(u) u^2\\
2 G(u) u^2&&-\mathcal{L}-G(u)(1-u^2)  \\
\end{bmatrix},
\end{equation}
with $\mathcal{L} \equiv \partial_{xx}-k_y^2+\gamma$ and $G(u) \equiv (1-u^2)^{-3}$. 
In Fig.~\ref{fig:trinst}, we show the pseudo-color map of the TI gain $g_{TI}=Re(\lambda)$ versus $u_0$ and $k_y$. As shown, the TI peak gain is relatively large only around the transition region (see also inset (a)). It rapidly drops for $u_0>1.5$ since saturation weakens the instability and also in the Kerr region $u_0<1$ where solitons become shallow (in Kerr limit, the TI gain is proportional to soliton peak intensity). For the observations reported in Fig.~\ref{fig:exp}, TI does not build up because the characteristic gain length $Z_g = 8 Z_{nl}/g_{TI}$ (assuming a relevant amplification factor of fluctuations $\sim \exp(8)$) substantially exceeds the crystal length $L_z=2.5$ mm, i.e. TI remains long-range. Nonetheless, the onset of the TI turns out to be strongly dependent on actual beam size used, and can be made to dominate phenomenology using tighter beams. In fact, when narrowing the FWHM by a factor $f$, a soliton with fixed $u_0$ is expected to form with a bias voltage higher by the same factor $f$, but the gain length with be reduced by a factor $f^2$, this due to the quadratic nature of the electro-optic effect. Indeed, by reducing the FWHM down to 10 $\mu$m, as shown in the example of Fig.~\ref{fig:trinst} (b, c) for $u_0=0.21$ (deep in the Kerr limit), the process of raising the voltage to balance the enhanced diffraction (inset (b)), leads to an observable break-up in $y$ (inset (c)).
In particular, Fig.~\ref{fig:trinst}(c), obtained with bias $V=705$ V (i.e., the same bias that yields stable soliton formation for wider input beam with FWHM of 12.7 $\mu$m), shows that TI already becomes observable when diffraction is under-compensated by the nonlinear response. The observed average period of the breakup $\Delta Y \simeq 20 \mu$m is found to correspond to $k_y=2\pi X_0/\Delta Y=0.46$, which is larger than the calculated most unstable wavenumber $k_y=0.24$.  We attribute this mismatch  to the fact that, experimentally, the onset of TI occurs dynamically before the soliton condition is reached in terms of bias. In practice, the onset of TI poses a limitation to the minimum physical width that allows for the observation of unbroken soliton stripes. Evidently, since TI is a product of a low-dimensional nonlinear wave forming inside a higher-dimensional nonlinear host, we do not expect it to play a role for 2+1D solitons even in the Kerr-limit, a regime that has yet to be investigated.

{\bf Conclusions.} We have reported the first observation of 1+1D photorefractive stripe solitons that obey the paradigm Kerr nonlinear Schr\"odinger Equation (NLSE) that governs nonlinear waves across a wide variety of different systems. The Kerr regime is demonstrated comparing experimental soliton existence conditions to the theoretical screening soliton existence curve, here derived analytically for the first time. The achievement is made possible by the enhanced nonlinear response in photorefractive KLTN, that permits the exploration of the strongly unsaturated Kerr regime.  Findings connect an accessible experimental setup to a general theoretical paradigm, opening up a new arena for the study of nonlinear waves with relevance for different fields, also beyond optics. They also suggest new routes of study, such as the transition from integrable to nonintegrable waves and the role of extra dimensions \cite{Xin2022}, or even the interaction between integrable and nonintegrable solitons.
Most importantly, since the Kerr solitons are observed in a bulk environment, they may allow, in future developments, the experimental analysis of how dimensionality plays a role in hereto little understood and investigated 2+1D Kerr physics. 
\section*{Acknowledgments}
Funding from PRIN 2020 MUR Project No. 2020X4T57A is acknowledged. L. F. and E. D. also acknowledge support from PNRR MUR Project No. PE0000023-NQSTI, Sapienza-Ricerca di Ateneo Projects. S.T. also acknowledge European Union for grant  PRIN2022NCTCY - NextGenerationEU. A. J. A. acknowledge support from the State of Israel Ministry of science technology and Space, Grant No. 4698.
The authors also thank P. M. Santini, M. Onorato, C. Conti and D. Maestrini for extremely valuable discussions and/or suggestions on earlier manuscript versions.

\end{document}